\documentclass[prl,aps,twocolumn,superscriptaddress,showpacs]{revtex4}
\usepackage{amsmath,amssymb}
\usepackage[utf8x]{inputenc}
\usepackage[english]{babel}
\usepackage{graphicx,epsfig}
\usepackage{hyperref}
\usepackage{txfonts}

\usepackage[usenames,dvipsnames]{color}

\begin{document}

\title{Quantum statistical ensemble for emissive correlated systems}
\author{Alexey M. Shakirov}
\affiliation{Russian Quantum Center, Novaya street 100A, 143025 Skolkovo, Moscow Region, Russia}
\affiliation{Department of Physics, Lomonosov Moscow State University, Leninskie gory 1, 119992 Moscow, Russia}
\author{Yulia E. Shchadilova}
\email[Correspondence should be addressed to ]{yes@rqc.ru}
\affiliation{Russian Quantum Center, Novaya street 100A, 143025 Skolkovo, Moscow Region, Russia}
\author{Alexey N. Rubtsov}
\affiliation{Russian Quantum Center, Novaya street 100A, 143025 Skolkovo, Moscow Region, Russia}
\affiliation{Department of Physics, Lomonosov Moscow State University, Leninskie gory 1, 119992 Moscow, Russia}

\begin{abstract}
Relaxation dynamics of complex quantum systems with strong interactions towards the steady state is a fundamental problem in statistical mechanics.
The steady state of subsystems weakly interacting with their environment is described by the canonical ensemble which assumes the probability distribution for energy to be of the Boltzmann form.
The emergence of this probability distribution is ensured by the detailed balance of the transitions induced by the interaction with the environment.
Here we consider relaxation of an open correlated quantum system brought into contact with a reservoir in the vacuum state.
We refer to such a system as emissive since particles irreversibly evaporate into the vacuum.
The steady state of the system is a statistical mixture of the stable eigenstates arising due to the binding energy.
We found that, despite the absence of the detailed balance, the stationary probability distribution over these eigenstates is of the Boltzmann form in each $N$-particle sector.
A quantum statistical ensemble corresponding to the steady state is characterized by different temperatures in the different sectors, in a contrast to the Gibbs ensemble.
We investigate the transition rates between the eigenstates to understand the emergence of the Boltzmann distribution {and find their exponential dependence on the transition energy.
We argue that this property of transition rates is generic for a wide class of emissive quantum many-body systems.}
\end{abstract}

\maketitle

Whether and how a quantum system brought out of equilibrium reaches its steady state is a fundamental question which has recently attracted much attention \cite{Hofferberth2007,Polkovnikov2011,Trotzky2012,Gring2012,Langen2013,Eisert2015}.
Expectation values of local observables after relaxation are typically determined by integrals of motion while the memory of microscopic details of an initial state is lost.
In quantum systems the mechanism of thermalization is rooted in the properties of individual eigenstates as stated in the eigenstate thermalization hypothesis (ETH) \cite{Deutsch1991,Srednicki1994,Rigol2008}.
The basic statement of ETH for nonintegrable systems is the smoothness of eigenstate expectation values of local observables as functions of eigenenergies.
For systems with integrals of motion this statement remains true if eigenstates are taken from the same subspace \cite{Rigol2007,Cassidy2011,Calabrese2011}.
In the steady state the density matrix of any subsystem is diagonal and its elements are the same for the typical initial eigenstates of the full system \cite{Popescu2006,Linden2009,Genway2010,Genway2012}.
ETH may break down for rare eigenstates in the low-energy part of the spectrum \cite{Biroli2010,Roux2010}.

A subsystem weakly coupled to the rest of the system can be viewed as an open system with its surroundings acting as a reservoir.
Transitions induced by the coupling to the reservoir govern relaxation dynamics of the open system to the steady state which is described by the (generalized) Gibbs ensemble \cite{Goldstein2006,Eckstein2008,Znidaric2010,Riera2012}.
These transitions satisfy the detailed balance principle in the steady state and the probability distribution over many-body eigenstates of the open system is of the Boltzmann form.
However, this general scenario does not describe a special case of an open correlated quantum system brought into contact with a reservoir in the vacuum state.
{In this case particles evaporating from the system never return, and therefore the detailed balance principle does not hold.
In the following we refer to such a system as emissive.}
Although particles can only leave the system, its steady state may be populated if there is a binding energy for escaping particles.
Such a populated steady state is a statistical mixture of multiple eigenstates whose probability distribution is not a priori known.
The quantum statistical ensemble for emissive systems defined in this way has never been considered so far.

\begin{figure}[t]
\centering
\includegraphics[width=1\columnwidth]{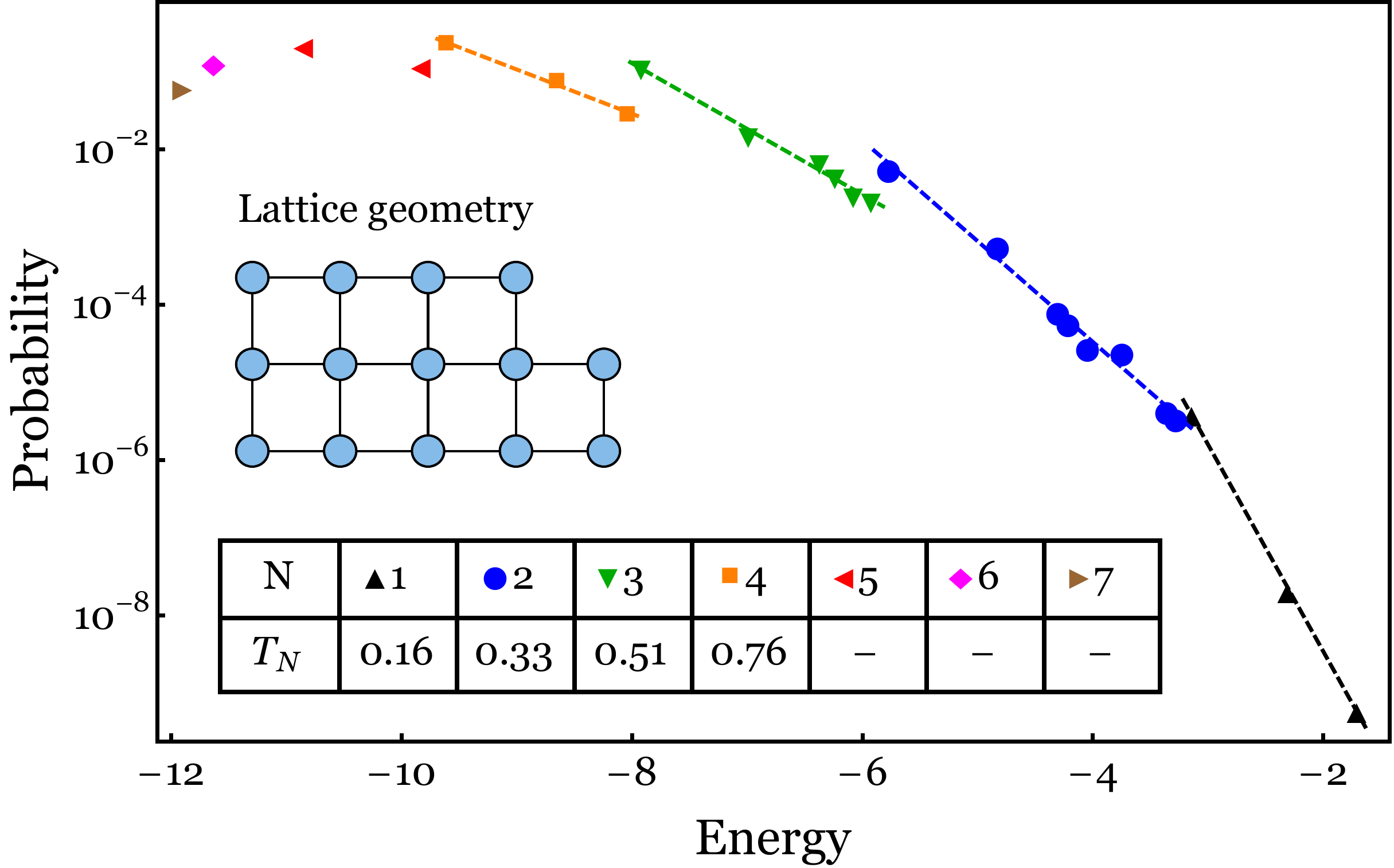}
\caption{
The stationary probability distribution of the emissive system of hard-core bosons on the $14$-site lattice (geometry is shown in the inset) is shown.
The initial state of the system is the maximally occupied pure state with $N_{0}=14$.
In sectors with at least $3$ stable states the data is fitted with the Boltzmann distribution $P^{N}_{n}\propto\exp(-E_{n}/T_{N})$.
The temperatures $T_{N}$ are presented in the table.}
\label{fgr:stationary}
\end{figure}

Open systems coupled to a vacuum reservoir appear ubiquitously in various fields of natural sciences including surface science \cite{Cooks2006, Somorjai2010}, quantum optics \cite{Scully1997,Walls2008}, nuclear physics \cite{Taleyarkhan2002} and astrophysics \cite{Spitzer1987}.
In the field of cold atoms an established experimental example is a trapped atomic or molecular gas in a vacuum chamber \cite{Mewes1997,Bloch2005,Bloch1999}.
Collisions of trapped particles result in internal equilibrium of the gas which alters as particles escape.
The loss of particles can be accompanied by cooling.
This process of evaporative cooling is a key technological development used to achieve Bose-Einstein condensation of cold atoms in magneto-optical traps \cite{Hess1986,Davis1995,Ketterle1996,McKay2010}.
The systems studied in experiments with ultracold atoms typically contain from $10^{3}$ to $10^{6}$ particles.
Statistical properties of these systems are usually probed by measuring their local observables.
Recent experimental progress has made it possible to study smaller systems \cite{Bakr2009,Serwane2011,Zimmermann2011,Bourgain2013}.
For these systems a direct measurement of the probability distribution over many-body states can be realized.
In particular, it would allow to probe statistical properties of the steady state of emissive quantum systems.

In this paper we study an emissive quantum system of hard-core bosons on a lattice.
We demonstrate that the probability distribution in the steady state is of the Boltzmann form in each $N$-particle sector.
We characterize this steady state by a quantum statistical ensemble and discuss its application to calculating expectation values of observables.
We show that the physical mechanism behind the emergence of the Boltzmann distribution is rooted in the behaviour of the transition rate.
This rate appears to be a regular function of the transition energy though there is no detailed balance with the reservoir.
We connect this behaviour with statistical properties of off-diagonal matrix elements of local annihilation operators entering expressions for transition rates.

We study a system of hard-core bosons on a two-dimensional lattice coupled to the vacuum reservoir with the Hamiltonian $H=H_{S}+H_{R}+H_{I}$.
The lattice is described by the Hamiltonian
\begin{equation}
H_{S}=-\sum_{\langle ij\rangle}h_{ij}(b^{\dag}_{i}b_{j}+b^{\dag}_{j}b_{i})
\label{eqn:Hamiltonian}
\end{equation}
with a constraint that each site can be occupied with no more than one particle.
Here $b^{\dag}_{i}$ ($b_{i}$) is a creation (annihilation) operator on the site $i$, $h_{ij}$ are hopping amplitudes and the sum runs over the pairs of the nearest-neighbour sites.
For calculations we use a lattice of $14$ sites (see inset of Figure 1).
To ensure non-integrablility of the lattice, the different values are assigned to the hopping amplitudes $h_{ij}$.
These values are taken from the interval between $0.8$ and $1.2$ (in arbitrary units of energy).
The vacuum reservoir is described by the Hamiltonian $H_{R}=\sum_{k}\varepsilon_{k}a^{\dag}_{k}a_{k}$ with operators $a^{\dag}_{k}$ ($a_{k}$) creating (annihilating) particles in reservoir modes $k$ with energy $\varepsilon_{k}$.
The interaction between the lattice and the reservoir is introduced through the  Hamiltonian $H_{I}=\alpha\sum_{ki}(a^{\dag}_{k}b_{i}+b^{\dag}_{i}a_{k})\theta(\varepsilon_{k}-\varepsilon_{0})$,
where the unit step function $\theta(\varepsilon_{k}-\varepsilon_{0})$ accounts for the binding energy $\varepsilon_{0}$.
This binding energy is a tunable parameter in our calculations.
We consider the weak coupling limit $\alpha\ll1$ and assume that escaping particles immediately lose the coherence with the lattice and do not return back to it.

We describe the reduced dynamics of the lattice by a master equation \cite{Breuer2002,Gardiner2004} for its density matrix $\rho_{S}$.
The density matrix is represented in the basis of eigenstates of $H_{S}$.
These eigenstates $|N,n\rangle$ are divided into sectors according to the number of particles $N$ in the lattice.
Corresponding eigenenergies are denoted by $E^{N}_{n}$.
We perform leading order calculations of the transitions between the eigenstates generated by the coupling to the reservoir \cite{Shakirov2015}.
In the absence of special symmetries of the lattice there are two constraints on these transitions: (i) $\Delta N=-1$ which is due to the type of coupling, and (ii) $\Delta E<-\varepsilon_{0}$ which is due to the binding energy for escaping particles.

We consider the dynamics of the system at the coarse-grained time scale $dt\sim\tau_{loss}$, where $\tau_{loss}$ is the characteristic time of the particle loss process.
In the weak coupling regime $\tau_{loss}\propto\alpha^{-2}$ is much larger than the dephasing time, so that only the diagonal elements of the density matrix contribute to dynamics.
These elements $\langle N,n|\rho_{S}|N,n\rangle\equiv P^{N}_{n}$ may be viewed as the probability distribution of the lattice over its eigenstates.
The master equation is reduced to the system of the rate equations for this probability distribution
\begin{equation}
\dfrac{d}{dt}P^{N}_{n}=\sum_{m}R^{N+1}_{nm}P^{N+1}_{m}-
\sum_{m}R^{N}_{mn}P^{N}_{n},
\label{eqn:Master}
\end{equation}
where $R^{N}_{mn}$ is the transition rate from the state $|N,n\rangle$ into the state $|N-1,m\rangle$.
For transitions allowed by the selection rules the rates are estimated using Fermi's golden rule as
\begin{equation}
R^{N}_{mn}=2\pi\Omega_{0}\alpha^{2}\sum_{i=1}^{L}|\langle N-1,m|b_{i}|N,n\rangle|^{2}.
\label{eqn:Rates}
\end{equation}
Here the step-like density of states of the reservoir $\Omega=\Omega_{0}\theta(\varepsilon-\varepsilon_{0})$ has been used {that corresponds to the two-dimensional motion of escaping particles}.
The particle loss process can be viewed as a sequence of transitions which brings the system from the initial to one of the stable states from which no further transitions can occur (see Supplementary).
There is the trivial stable state with no particles and occupied stable states which appear due to the constraint on the transition energy.
For calculations we use the maximally occupied pure state with $N_{0}=14$ as the initial one and choose $\varepsilon_{0}=0$ to maximize the number of the stable states which can be achieved (see Supplementary).
For the $14$-site lattice this number is $26$ and the maximal occupation of the stable states is $7$.

The probability distribution of the system in a steady state is shown in Figure 1.
In each $N$-particle sector of the Hilbert space (having at least $3$ stable states) this stationary distribution is found to be of the Boltzmann form.
{Parameters of the exponential fit of the data \footnote{The coefficient of determination for a linear fit of the numerical data in the semi-logarithmic scale is around 0.99.
We observe the exponential behavior in the range of more than two orders of magnitude that certainly allows us to distinguish this dependence from e.g. the power law function.}
\begin{equation}
P^{N}_{n}=\frac{P_{N}}{Z_{N}}e^{-E^{N}_{n}/T_{N}}.
\label{eqn:Ensemble}
\end{equation}
define a quantum statistical ensemble.
This ensemble is characterized by the set of temperatures $T_{N}$ and the probabilities $P_{N}$ for the system to have $N$ particles in the steady state.
The partial partition functions $Z_{N}=\sum_{n}e^{-E^{N}_{n}/T_{N}}$ are summed over achievable stable states.
We assign zero probabilities to the unstable states which are not present in the ensemble.}
Stationary expectation values of the observables are given by a standard expression $\overline{\langle A\rangle}=\sum_{Nn}P^{N}_{n}\langle N,n|A|N,n\rangle$.
We note that the steady state can be characterized by this statistical ensemble for an arbitrary initial state of the system (see Supplementary).
Variation of the initial state only changes parameters $P_{N}$ and $T_{N}$ of the distribution.

\paragraph{Emergence of the distribution.}
The emergence of the Boltzmann distribution in the emissive system can be understood from statistics of the transitions which the system undergoes as particles evaporate.
For each sector we calculate the occurrence probabilities of the intermediate states, i.e. the states through which the system evolves from the initial to one of the stable states.
These distributions in sectors (referred to as intermediate) are shown in Figure 2 as functions of the energy of the states.
In sectors with $N=3\div11$ the dependence is smooth and fluctuations of the data are less than the size of points.
In each sector we consider the part of the spectrum accumulating the probability of $0.8P_{N}$ and approximate corresponding points by the Boltzmann distribution (\ref{eqn:Ensemble}).
For $N\leqslant7$ the data includes stable states and the temperatures are close to those in Figure 1.
We note that measuring the occurence probabilities of the intermediate states will require collecting statistics from an ensemble of emissive systems.
Assuming that the spectrum of the system is discrete and not degenerate, a sequence of the intermediate states can be directly determined by measuring the energies of escaping particles.

\begin{figure}[t]
\begin{flushleft}
\includegraphics[width=1\columnwidth]{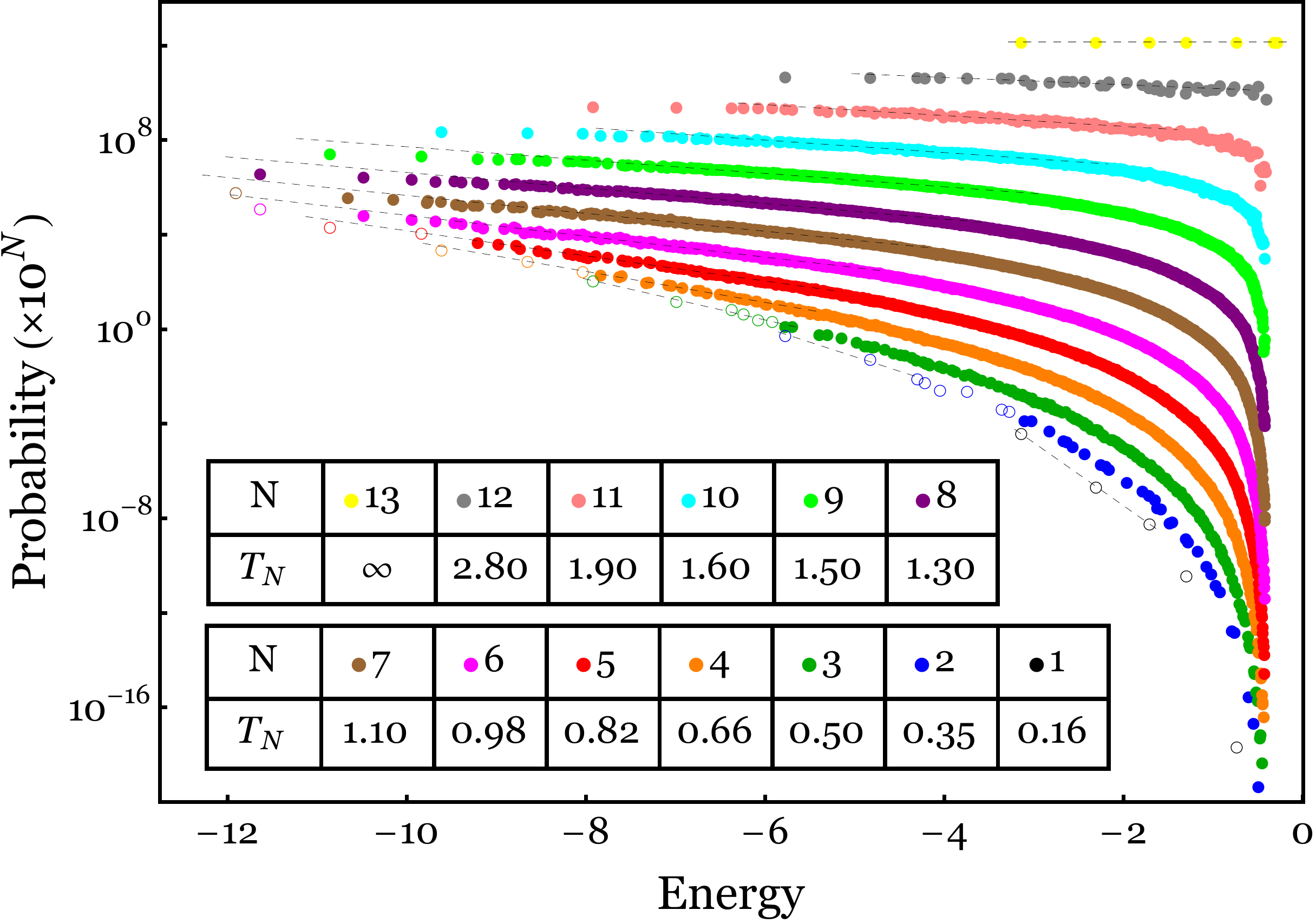}
\end{flushleft}
\caption{
The probability distribution of the intermediate states in each $N$-particle sector is shown. 
The initial state of the system is the maximally occupied pure state with $N_{0}=14$.
For illustrative purposes the rates are multiplied by the factor $10^{N}$.
The stable states are indicated with empty circles.
In each sector the points accumulating $80\%$ of the total probability are fitted with the exponential functions $P_{N}(E)\propto\exp(-E/T_{N})$.
The temperatures $T_{N}$ are presented in the table.}
\label{fgr:intermediate}
\end{figure}

The smooth energy dependence of intermediate probability distributions can be understood from the statistical properties of the transition rates in the system.
Figure 3 shows $R^{N}_{mn}$ plotted as a function of the transition energy $\varepsilon=E_{n}-E_{m}$.
The data for $R^{N}(\varepsilon)$ in each sector can be approximated by a regular dependence with state-to-state fluctuations.
For an arbitrary initial $N$-particle ensemble the probability distribution in $(N-1)$-particle sector after the loss of a particle is determined by
\begin{equation}
P^{N-1}_{m}\propto\sum_{n}R^{N}_{mn}P^{N}_{n}.
\label{eqn:Convolution}
\end{equation}
Possible state-to-state fluctuations of $P^{N}_{n}$ are convolved with the regular function of the transition energy and almost do not translate into the probability distribution in the following sector.
The probability distribution becomes a smooth function of energy after the loss of several particles if transition rates in neighboring sectors are uncorrelated.

The analysis of lattices of different sizes (see Supplementary) shows that the transition rate is always a regular function of the transition energy with state-to-state fluctuations.
The fluctuations decrease with the increase of the size of the lattice.
We expect that they vanish in the thermodynamic limit, though it is not evident from the available range of lattice sizes.
{This assumption is supported by the analysis of the transition rates in the thermodynamic limit presented below.}

\begin{figure}[t]
\begin{flushleft}
\includegraphics[width=1\columnwidth]{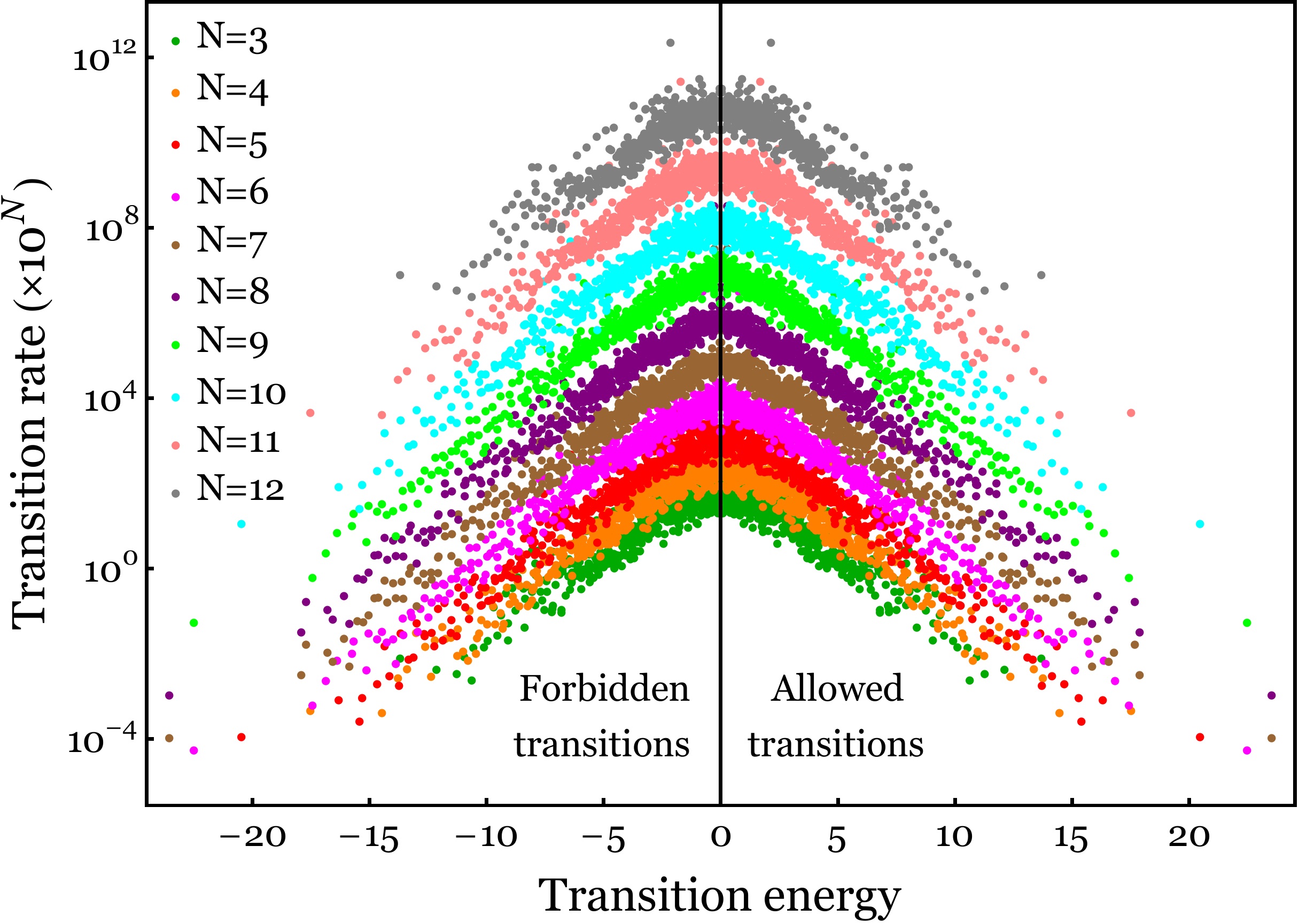}
\end{flushleft}
\caption{
The transition rates $R^{N}_{nm}$ from the states in the $N$-particle sector to the states in the $(N-1)$-particle sector as a function of the transition energy $\varepsilon=E^{N}_{n}-E^{N-1}_{m}$ for $N=3,\ldots,12$.
The allowed transitions correspond to $\varepsilon>\varepsilon_{0}=0$.
For illustrative purposes $900$ points are shown for each $N$ and the rates are multipled by the factor $10^{N}$.}
\label{fgr:transitions}
\end{figure}

{
\paragraph{Thermodynamic limit.}
Let us show that the exponential dependence of the transition rate on the transition energy is a universal feature of quantum many-body systems coupled to the vacuum reservoir.
For this purpose we consider evaporation in a generic system of $N\gg1$ particles with well defined energy $E$.
Temperature $T$ and chemical potential $\mu$ of the system are related to the density of many-body states $A(N,E)$ via well known thermodynamic relations $1/T=\frac{\partial}{\partial E}\ln A$, $\mu=-T\frac{\partial}{\partial N}\ln A$.
It is reasonable to assume that the energy $\varepsilon$ of a particle escaping to the vacuum obeys the Boltzmann distribution with some temperature $T'$ (not necessarily equal to $T$, e.g. because of the adiabatic cooling during evaporation). 
The probability density of the energy of the leaving particle is $p(\varepsilon)\propto a(\varepsilon)e^{-\varepsilon/T'}$, where $a(\varepsilon)$ is the density of single particle states in the vacuum reservoir.
On the other hand, the same probability can be estimated with the Fermi's golden rule as $p(\varepsilon)\propto R^{N}(\varepsilon)A(N-1,E-\varepsilon)$.
One can use the first order Taylor series expansion of $\ln A$ and the thermodynamic relations above to show that $A(N-1,E-\varepsilon)=A(N,E)e^{-(\varepsilon-\mu)/T}$.
Comparing two expressions for $p(\varepsilon)$, we conclude that $\ln R(\varepsilon)=-(1/T'-1/T)\varepsilon-\ln a(\varepsilon)$.
For our specific case of two-dimensional reservoir with $a(\varepsilon>0)=\mbox{const}$ the transition rate is a regular exponential function of the transition energy, in accordance with results shown in Figure \ref{fgr:transitions}. 
The difference between the inverse temperatures of the gas and the leaving particle is determined by the slope angle of $\ln R(\varepsilon)$ function.
We conclude that an exponential dependence of the transition rate on the transition energy is a property of the wide class of emissive quantum many-body systems where escaping particles satisfy the Boltzmann distribution.}

The regular dependence of the transition rate on the transition energy relates to statistical properties of the off-diagonal matrix elements $\langle N-1,m|b_{i}|N,n\rangle$ which enter the expression (\ref{eqn:Rates}).
It is important to note that $b_{i}$ are non-Hermitian annihilation operators and the matrix elements are calculated between states in the different sectors, in contrast to known studies \cite{Rigol2009,Beugeling2015}.
We argue that the smooth dependence of these matrix elements on the transition energy is similar to that of eigenstate expectation values in ETH.
Fluctuations of the expectation values vanish as the size of the system increases \cite{Ikeda2013,Steinigeweg2013,Beugeling2014}.
From our scaling analysis we expect that fluctuations of $\langle N-1,m|b_{i}|N,n\rangle$ have the same property.

We note that an experimental verification of our findings will require measuring the probability distribution of a system under study over its many-body states \cite{Liphardt2002}.
This type of measurements would extend the method of determining temperature from average values of (local) observables.
While the latter is available both for finite-size and macroscopic systems, the former is only realizable for the systems of finite size.

\begin{figure}[t]
\centering
\includegraphics[width=1\columnwidth]{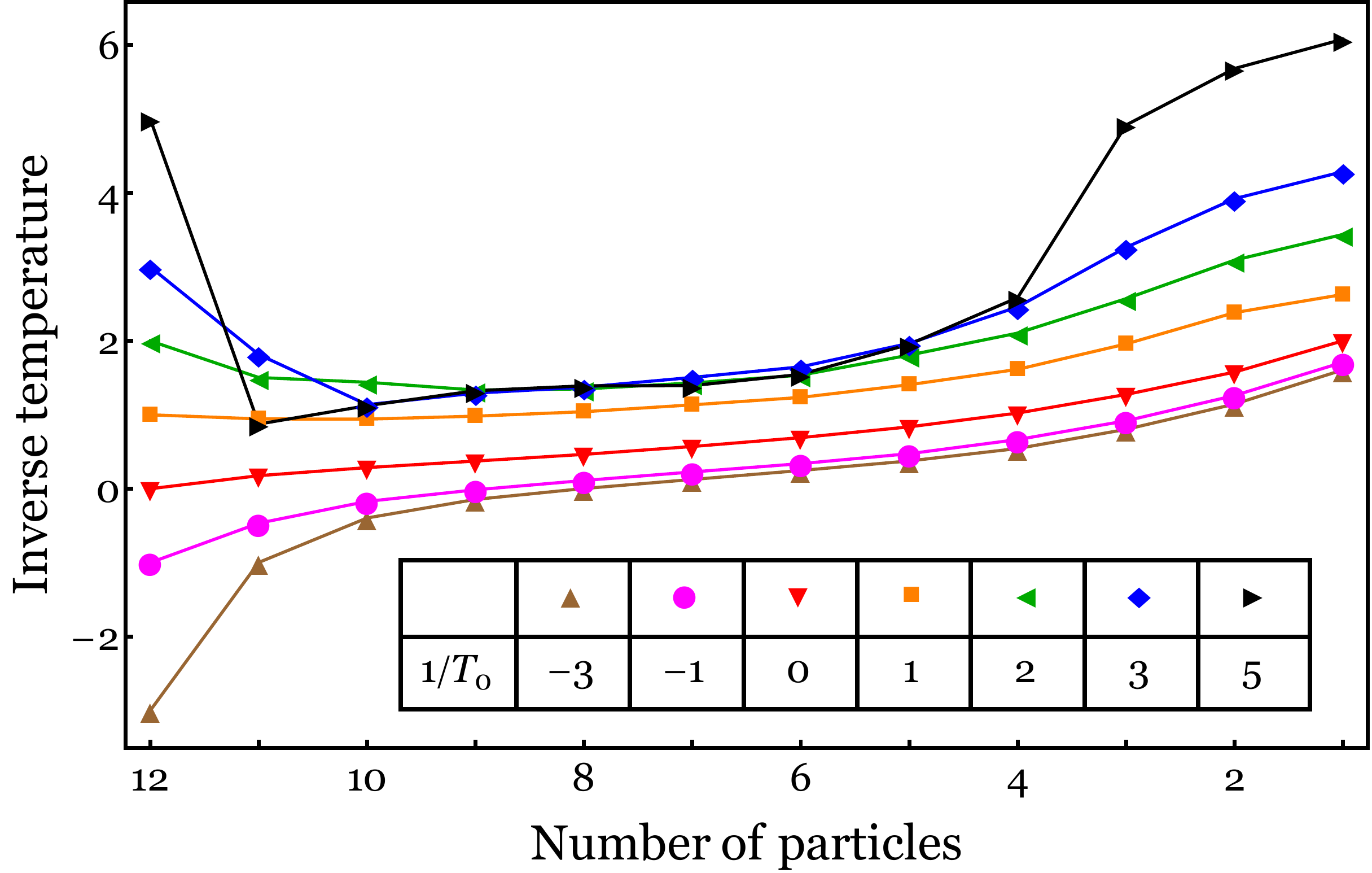}
\caption{
The plot shows the dependence of the inverse temperatures $1/T_{N}$ on the number of particles for different initial conditions which are defined by the set of initial inverse temperatures  $1/T_{0}$ (shown in the inset table). The plot demonstrates that system may undergo both heating and cooling process during the evaporation depending on the initial temperature of the system and the filling of the system. }
\label{fgr:4}
\end{figure}

\paragraph{Cooling/heating effect.}
Emissive systems are typically cooled down by the particle loss process since leaving particles accumulate on average more energy than the remaining ones.
Preparation of the system in a Gibbs state with $N_0<14$ allows us to study the cooling/heating effect in the system and its dependence on the initial temperature $T_{0}$.
Fig. \ref{fgr:4} shows the inverse temperature $1/T$ as a function of $N$ for the Gibbs states with $N_0=12$ and different initial temperatures.
The energy spectrum of the system is bounded both from below and above, so the initial states with negative $T_{0}$ can also be considered.
In the case of a negative initial temperature the system always cools down as particles escape.
In the case of a small positive initial temperature the loss of the first particle leads to the abrupt heating of the system.
This can be explained by the shift between the spectra of the neighbouring sectors which leads to the possibility of transitions from the lowest energy state in the $N$-particle sector to the excited $(N-1)$-particle states.
For the system under study this condition can be satisfied only for $N>7$ so that no heating is observed below half-filling.

\paragraph{Conclusions.}
To summarize, we demonstrate an emergence of the Boltzmann distribution in each $N$-particle sector of an emissive system of hard-core bosons on a lattice.
The steady state of the system is described by a quantum statistical ensemble characterized by a set of temperatures.
Smooth probability distributions are the consequence of the statistical properties of the transition rates in the system.
The transition rates are expressed in terms of the off-diagonal matrix elements of local annihilation operators and depend smoothly on the transition energy.
This allowed us to draw the analogy with ETH which asserts the smooth dependence of the eigenstate expectation values on the energy.
We expect that this feature of transition rates is a generic property of the correlated emissive quantum systems and the formation of the smooth Boltzmann distribution can be observed in contemporary experiments with ultracold atoms.
{The physics demonstrated in this paper is strongly determined by a particular choice of the reservoir which can only absorb particles.
This excludes thermalization within the same sector due to virtual processes and makes low-energy populated states stable.}

Authors are greatful to A. E. Antipov, P. V. Elyutin, P. Gri\v sins, B. Krippa, J. P. F. LeBlanc, and P. Ribeiro for useful discussions.
The authors acknowledge a financial support from the RFBR grant 14-02-01219 and the Dynasty foundation.

\newpage

\begin{widetext}

\begin{center}
{\large\bf Quantum statistical ensemble for emissive correlated systems\\
(supplementary materials)}
\end{center}

\begin{center}
Alexey M. Shakirov,$^{1,2}$ Yulia E. Shchadilova,$^1$ and Alexey N. Rubtsov$^{1,2}$
\end{center}

\begin{center} 
{\small\it$^1$ Russian Quantum Center, Novaya 100, 143025 Skolkovo, Moscow Region, Russia\\
\small\it$^2$ Department of Physics, Lomonosov Moscow State University, Leninskie gory 1, 119991 Moscow, Russia}\end{center}

\end{widetext}

\setcounter{figure}{0}   \renewcommand{\thefigure}{S\arabic{figure}}
\setcounter{equation}{0} \renewcommand{\theequation}{S.\arabic{equation}}
\setcounter{section}{0} \renewcommand{\thesection}{S.\Roman{section}}

\renewcommand{\thesubsection}{S.\Roman{section}.\Alph{subsection}}

\makeatletter
\renewcommand*{\p@subsection}{}  
\makeatother

\renewcommand{\thesubsubsection}{S.\Roman{section}.\Alph{subsection}-\arabic{subsubsection}}

\makeatletter
\renewcommand*{\p@subsubsection}{}
\makeatother

\section*{Hilbert space and transitions}

\begin{figure}[h!]
\centering
\includegraphics[width=0.9\columnwidth]{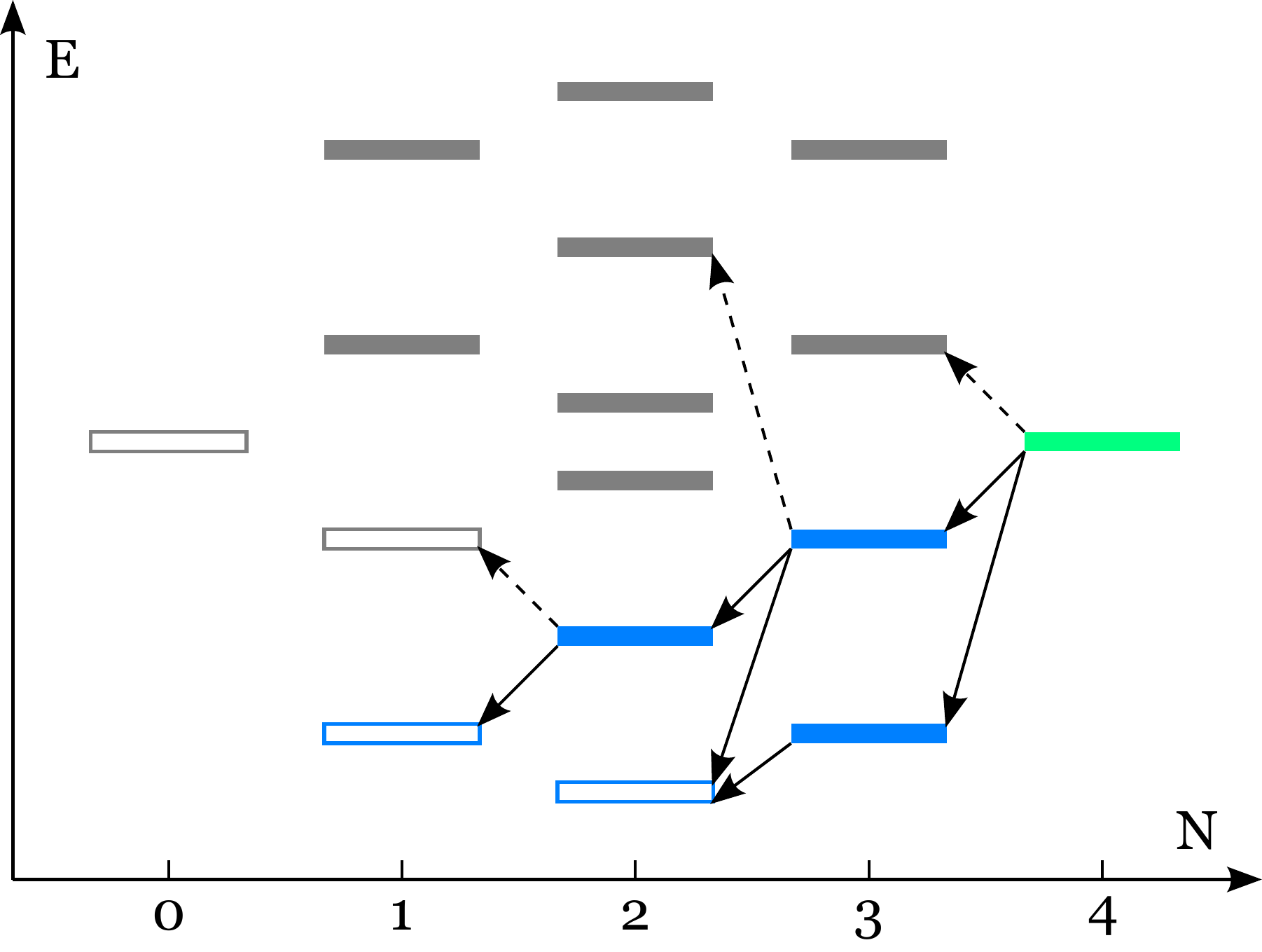}
\caption{\textbf{Graphical representation of the Hilbert space.}
The system of hard-core bosons on a lattice of $4$ sites is considered. Bars representing many-body eigenstates $|N,n\rangle$ are placed into the coordinate system 'number of particles - energy'.
Initial maximally occupied pure state is indicated by the green colour.
Allowed transitions and some forbidden transitions for $\varepsilon_{0}=0$ are shown with solid and dashed arrows correspondingly.
The achievable eigenstates are indicated with the blue colour.
The stable eigenstates are represented with empty bars.
After the loss of particles the systems in the ensemble can be in either the lowest energy $2$-particle eigenstate or the lowest energy $3$-particle eigenstate.}
\label{fgr:1}
\end{figure}

The Hilbert space of the system can be represented as the graph shown in Fig. \ref{fgr:1}.
The vertices of this graph represent the many-body eigenstates $|N,n\rangle$ which are classified by the number of particles $N$ and the energy $E_{n}$.
Directed edges connect these vertices according to the transitions between the eigenstates generated by the coupling to the vacuum reservoir.
The dynamics of the system can be considered as a walk through the graph.
Transitions should satisfy two constraints: (i) $\Delta N=-1$, (ii) $\Delta E<-\varepsilon_{0}$.
By possible transitions we mean all the transitions satisfying the first constraint.
We divide them into allowed transitions which also satisfy the second constraint and forbidden transitions which do not satisfy the second constraint.
We also introduce two categories for the eigenstates of the system.
For a given initial state we define an achievable eigenstate as an eigenstate which can be reached from the initial one through the allowed transitions.
We define stable eigenstates as the eigenstates from which no further transitions are allowed.
Only achievable stable eigenstates are present in the steady state of the system.
{We note that the dynamics of the system clearly does not satisfy the detailed balance principle since all possible transitions are irreversible.}

\section*{Binding energy}

\begin{figure}[h!]
\centering
\includegraphics[width=0.9\columnwidth]{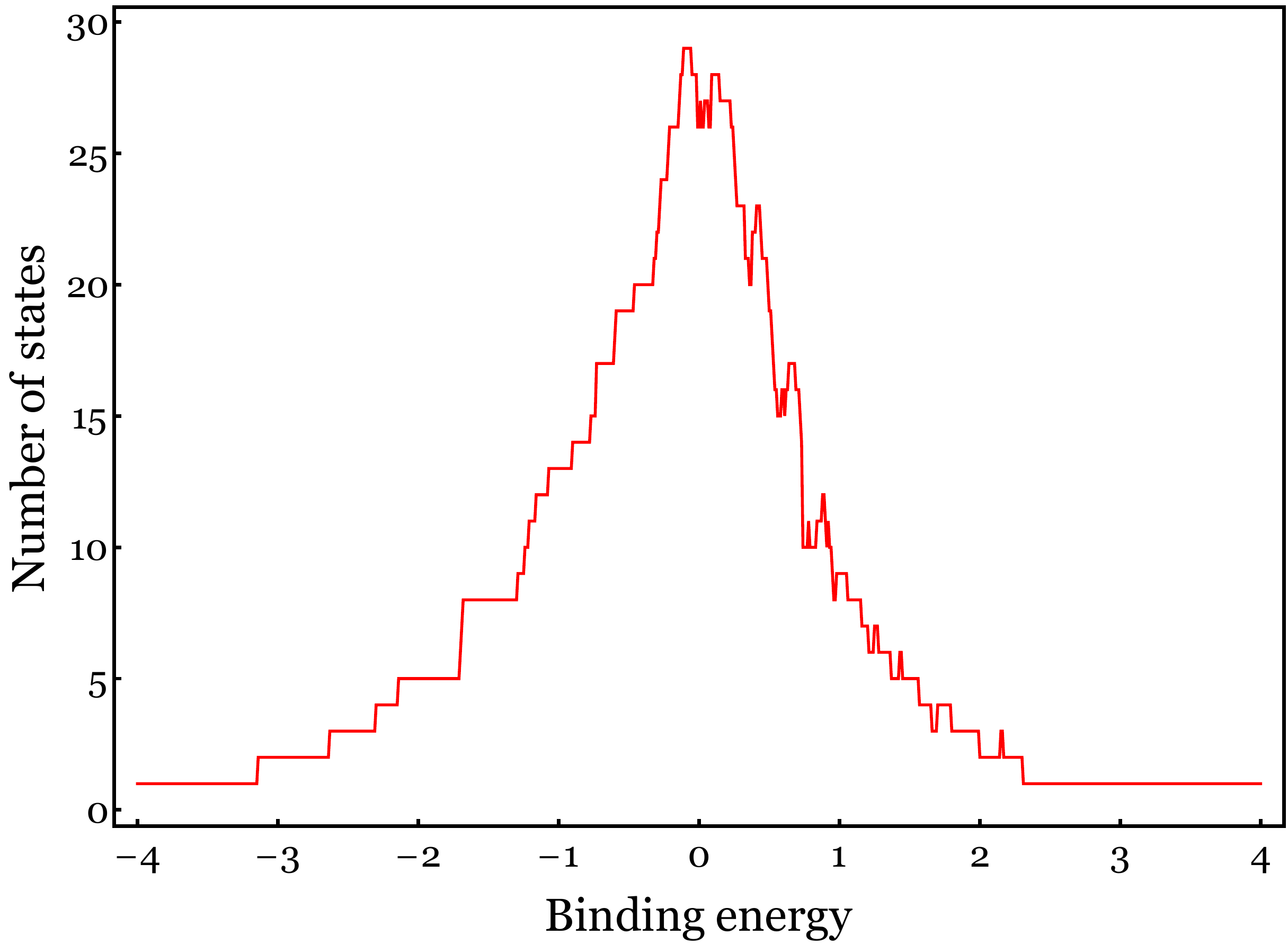}
\caption{\textbf{Number of achievable stable eigenstates.}
The system of hard-core bosons on a $14$-site lattice is prepared in the maximally occupied pure state.
The plot shows the number of achievable stable eigenstates as the function of the binding energy.}
\label{fgr:2}
\end{figure}

The binding energy $\varepsilon_{0}$ determines the allowed transitions and the eigenstates which are present in the steady state of the system.
The dependence of the number of these eigenstates on $\varepsilon_{0}$ for the initial maximally occupied pure state is shown in Fig. \ref{fgr:2}.
For the extreme values of the binding energy there is only one achievable stable eigenstate: (i) the empty eigenstate for $\varepsilon_{0}=-\infty$ (all possible transitions are allowed), (ii) the maximally occupied eigenstate for $\varepsilon_{0}=\infty$ (all possible transitions are forbidden).
For calculations we set $\varepsilon_{0} = 0$ which corresponds to the maximal value of the number of eigenstates present in the steady state of the system.

\newpage

\begin{figure}[t!]
\centering
\includegraphics[width=1\columnwidth]{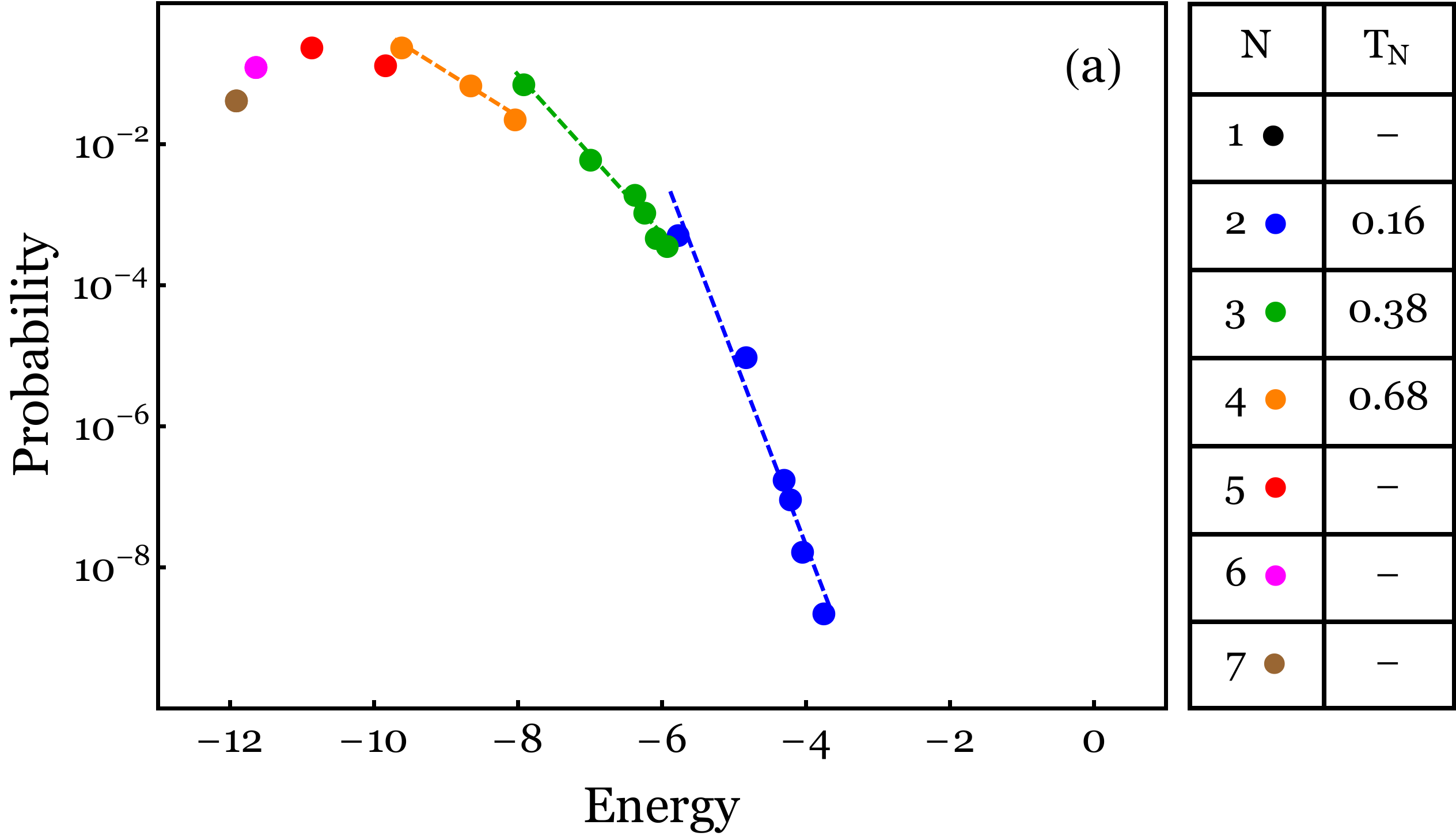}
\includegraphics[width=1\columnwidth]{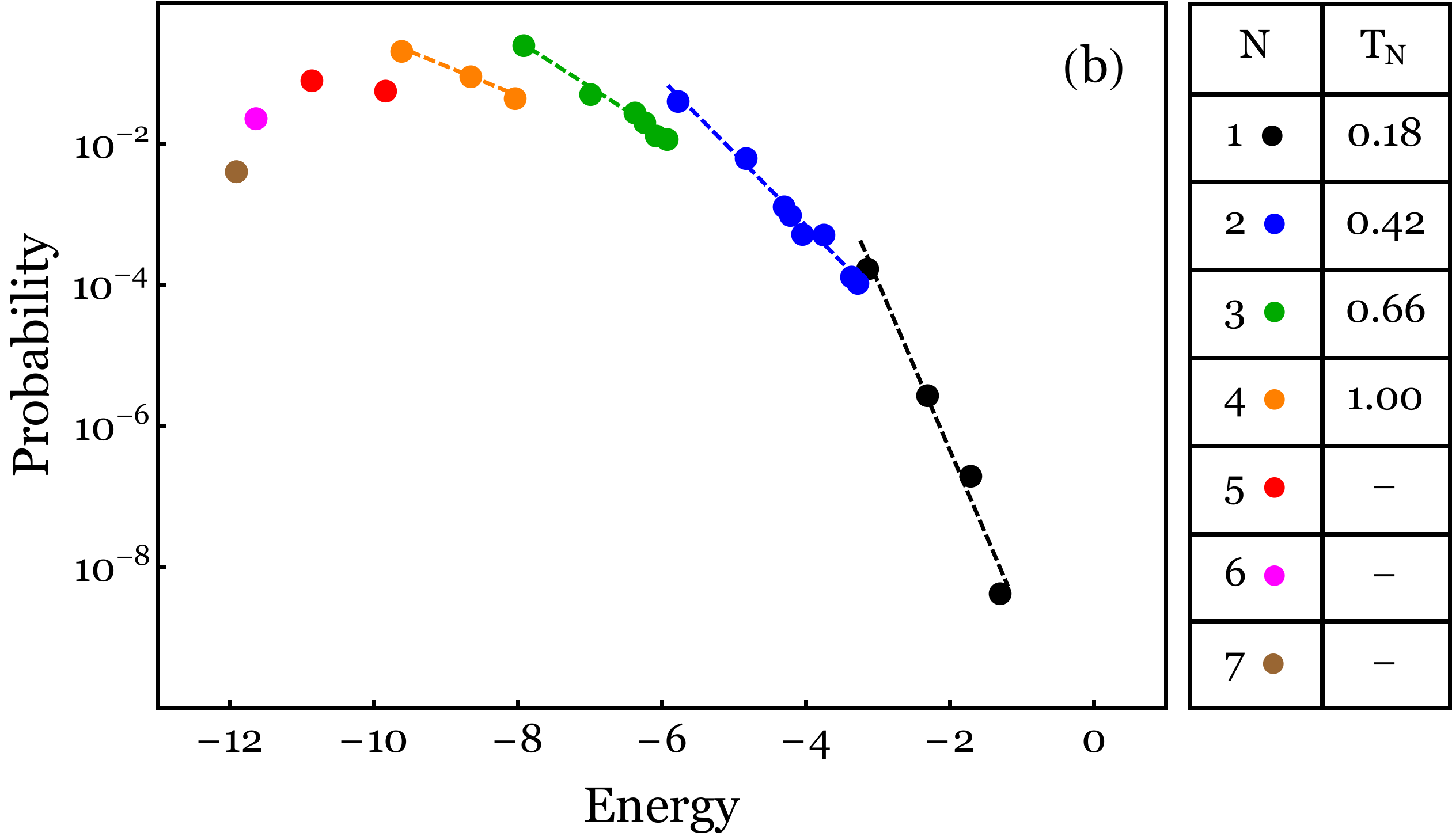}
\includegraphics[width=1\columnwidth]{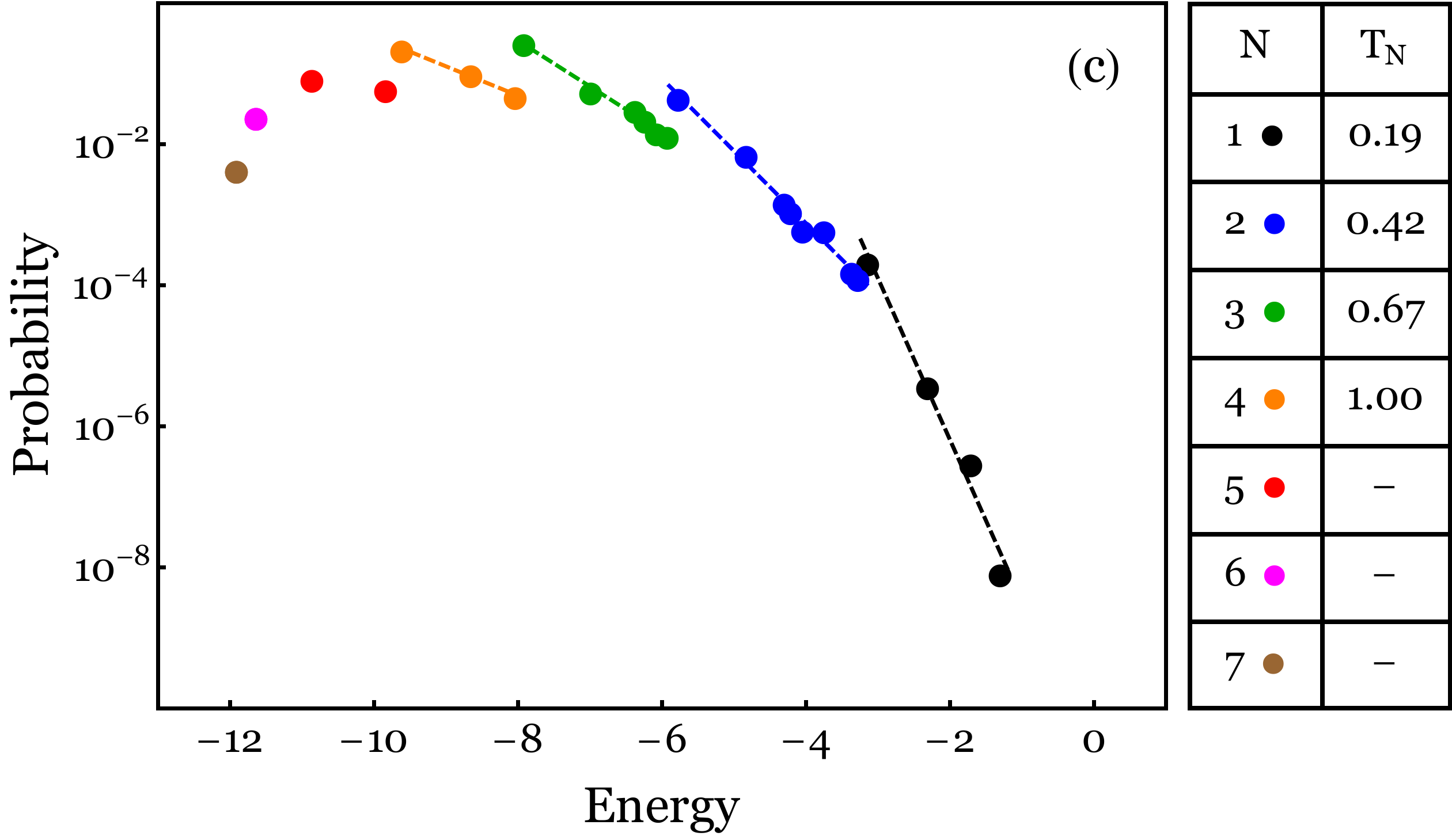}
\includegraphics[width=1\columnwidth]{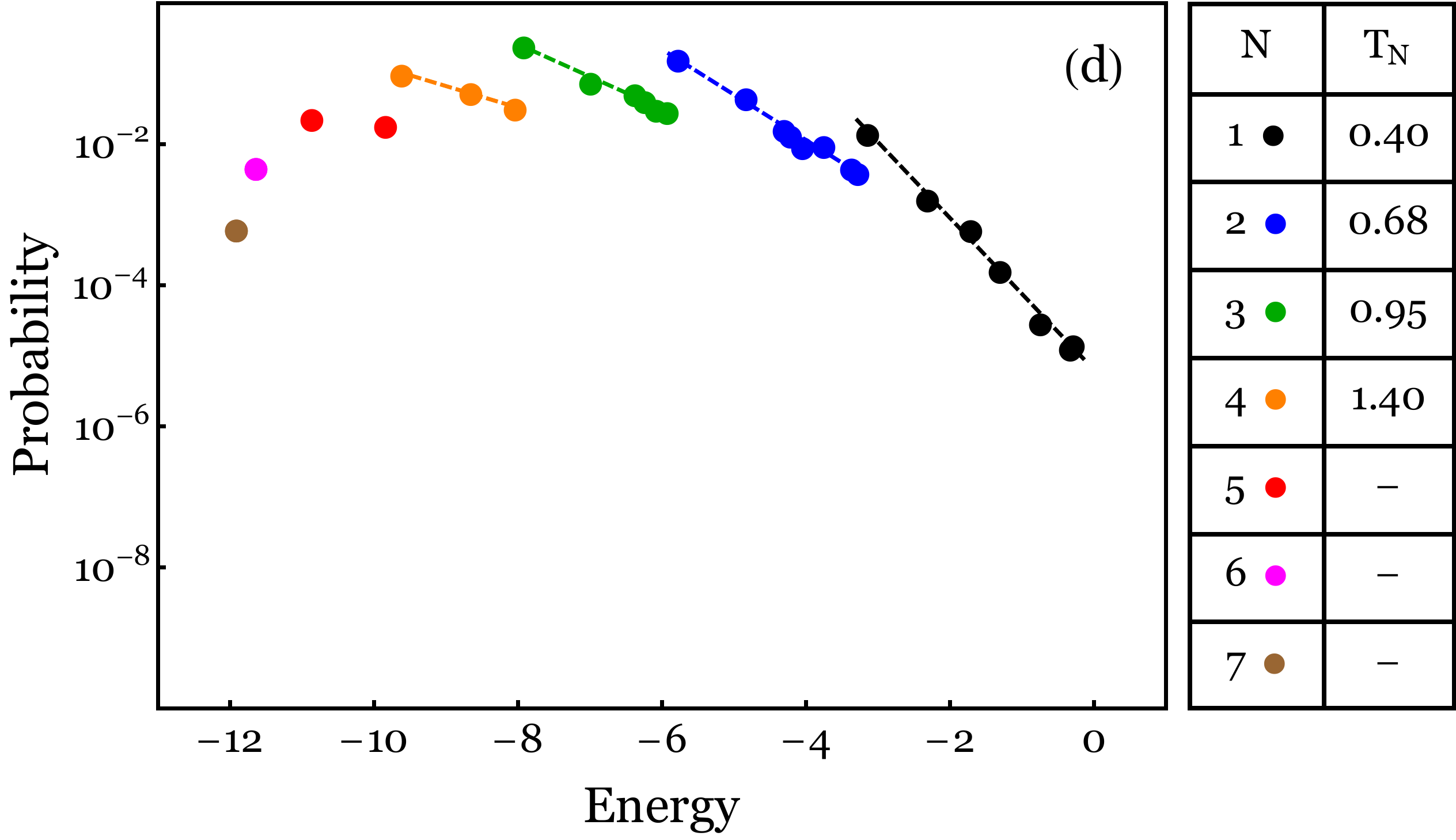}
\caption{\textbf{Stationary distributions.}
The system is prepared in pure states with $N_{0}=12$ particles and different energies: (a) $E_{0}=-3.00$, (b) $E_{0}=-0.06$, (c) $E_{0}=0.06$, (d) $E_{0}=3.00$. The stationary distributions are of the Boltzmann form in each $N$-particle sector for all these initial conditions (temperatures are presented in tables).}
\label{fgr:3}
\end{figure}

\section*{Dependence on the initial energy}

As stated in the paper, the steady state of the system is characterized by the Boltzmann distribution in each sector.
Here we demonstrate that the variation of the initial state of the system only changes parameters of the stationary distribution but not its shape.
Fig. \ref{fgr:3} shows stationary distributions for the initial pure states with $N_{0}=12$ particles and the different energies $E_{0}$.
We observe that (i) temperatures $T_{N}$ increase as $E_{0}$ increases, (ii) for the initial states with close energies stationary distributions are similar.

\section*{Scaling analysis}

\begin{figure}[h]
\centering
\includegraphics[width=1\columnwidth]{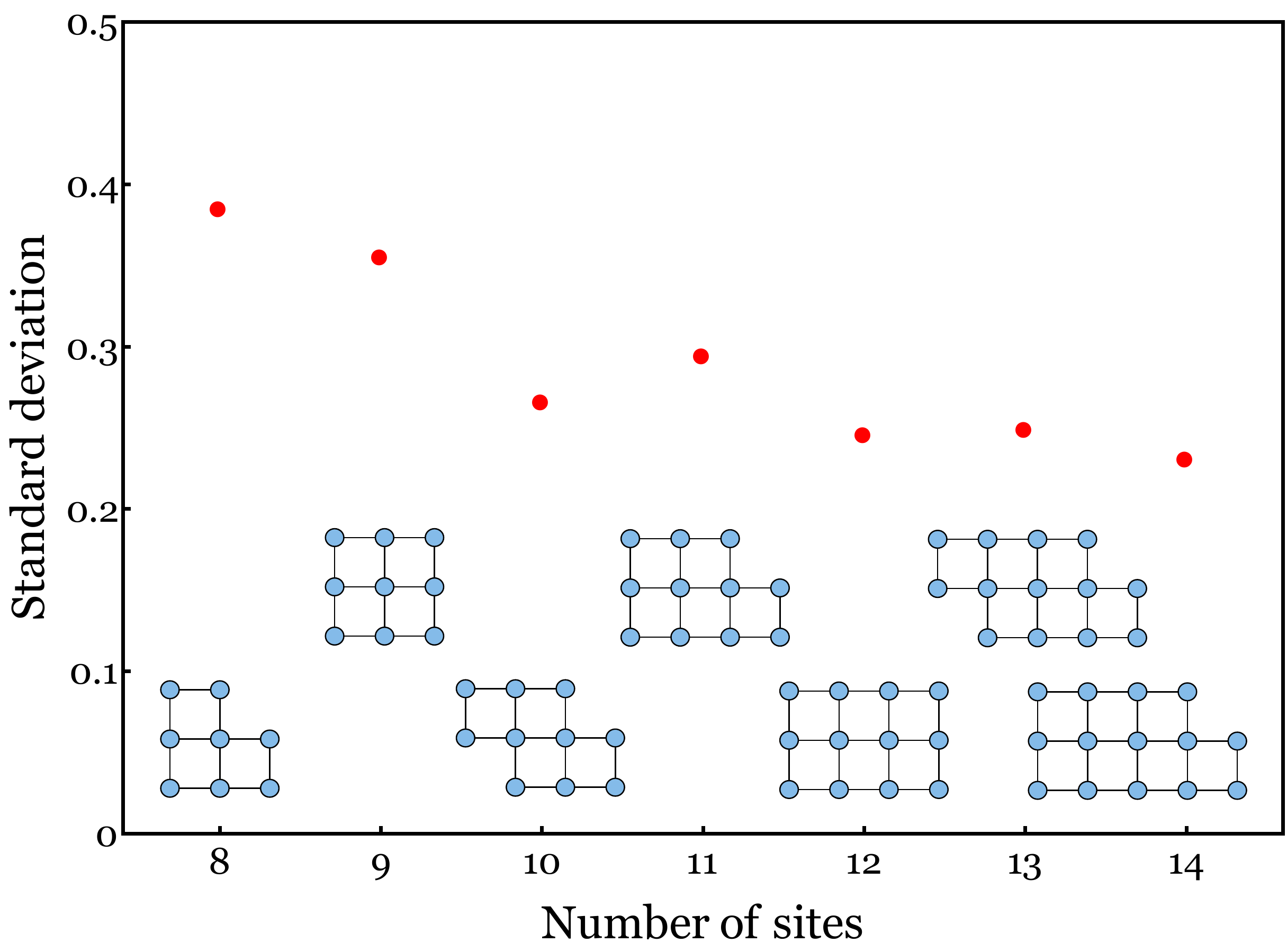}
\caption{\textbf{Scaling analysis.}
The plot shows the standard deviation of log$_{10}R$ in the interval of transition energies $[-0.1,0.1]$ as the function of the number of sites in the lattice.
Geometries of the lattices of different sizes are shown in the inset.}
\label{fgr:6}
\end{figure}

The amplitude of the state-to-state fluctuations of the transition rate as the function of the transition energy depends on the size of the system.
Fig. \ref{fgr:6} shows that the standard deviation of the state-to-state fluctuations becomes smaller as the size of the system increases.
Lattices of the different sizes are obtained by taking out sites from the 14-site lattice.

\end{document}